\documentclass[runningheads]{llncs}

\usepackage{amsmath,amssymb,amsfonts}
\usepackage{algorithmic}
\usepackage{graphicx}
\usepackage{textcomp}
\usepackage{xcolor}

\usepackage{acronym}
\usepackage{booktabs}
\usepackage{todonotes}
\usepackage{censor}
\usepackage{caption}
\usepackage{subcaption}

\acrodef{HDL}{Hardware Description Language}
\acrodef{DSL}{Domain-Specific Language}
\acrodef{CRV}{Constrained Random Verification}
\acrodef{API}{Application Programming Interface}
\acrodef{DSE}{Design Space Exploration}
\acrodef{ASIC}{Application-specific Integrated Circuit}
\acrodef{SoC}{System-on-Chip}
\acrodef{MAC}{multiply-accumulate}
\acrodef{HLS}{high-level synthesis}

\usepackage[backend=biber,sorting=none,doi=false,doi=false, url=true]{biblatex} 
\AtEveryBibitem{%
		\clearfield{urlyear}%
    \clearlist{location}
    \clearfield{publisher}
    \clearlist{publisher}
    \clearfield{pages}
} 

\usepackage[linesnumbered]{algorithm2e}

\usepackage{tikz, tikz-timing}
\usetikzlibrary{decorations, decorations.pathreplacing, calc, positioning, shapes, circuits.logic.US}
\usepackage{pgfplots, pgfplotstable}
\pgfplotsset{compat=1.18}

\usepackage[frozencache,cacheignoresfilecontents]{minted}

\newcommand{\mic}[1]{\mintinline{Haskell}|#1|}
\setminted[Haskell]{tabsize=2, breaklines, escapeinside=!!}
\usepackage{caption}
\usepackage{subcaption}

\usepackage{siunitx}

\usepackage{xspace}

\usepackage{csquotes}

\begin{document}
	
\title{Low-level optimizations in high-level HDLs:\\Is there a benefit?}

%
\author{
	Oliver Keszocze\inst{1}\orcidID{0000-0003-2033-6153} \and
	Tjark Petersen\inst{2}\orcidID{0000-0002-0239-511X} \and
	Arved Friedemann\inst{3}\orcidID{0000-0001-7252-264X} \and
	Matthias Bo Stuart\inst{2}\orcidID{0000-0003-0809-530X}
}
\institute{Computer Engineering, TU Clausthal,
	Clausthal-Zellerfeld, Germany\\ \email{oliver.keszoecze@tu-clausthal.de} 
	\and
	Embedded Systems Engineering, Technical University of Denmark,\\ Kongens Lyngby, Denmark\\
	\email{tjape@dtu.dk, mbst@dtu.dk}
	\and Independent Researcher\\ Hamburg, Germany\\ \email{arved.friedemann@stoppe.de}
}
\authorrunning{Keszocze, Petersen, Friedemann, and Stuart}

\maketitle

\begin{abstract}
This paper explores the applicability of functional programming to the design of \acp{ASIC}. 
We investigate the impact of designing \acp{ASIC} using high-level, abstract \ac{HDL} features versus employing low-level optimizations on the area of the synthesized circuits. The aim is to determine whether using low-level optimizations is beneficial and, if so, whether it is worth the added implementation effort. 

To carry out the investigation, we implement an unsigned bit-serial \ac{MAC} unit in 16 different configurations using the functional \ac{HDL} Clash. We make use of both low-level bit-manipulation techniques as well as Clash's high-level constructs for circuit design.

The experimental evaluation shows that some high-level constructs of Clash have negligible influence on the resulting circuit size, suggesting that using the full power of functional programming is a viable approach to hardware design. To evaluate the impact of the used \ac{HDL} itself, we also implemented versions of the MAC in Verilog. The experiments clearly show that there seems to be an inherent overhead in using Clash compared to Verilog code written by a seasoned engineer.
\end{abstract}

\acresetall

\section{Introduction and Related Work}

A major concern when using higher levels of abstraction for the description of circuits is that the abstractions add logic overhead to the circuit. Very little knowledge exists in the form of measurements or experiments that support or contradict this claim. This paper provides such knowledge by investigating the \acp{ASIC} implementation of a tiny but non-trivial coprocessor described in a high-level functional programming language with and without various low-level optimizations.

Many different modern \acp{HDL} have been proposed, many of which are either implemented as a framework or as a \ac{DSL} in an all-purpose programming language (e.g., Chisel and SpinalHDL for Scala~\cite{Bachrach2012, SpinalHDL} or MyHDL and Amaranth for Python~\cite{Decaluwe2004, amaranth}) or as an extension of an existing programming language (e.g., Clash~\cite{Baaij2010} for Haskell).\footnote{The interested reader is referred to the \texttt{awesome-hdl} GitHub repository at \url{https://github.com/drom/awesome-hdl} that hosts a curated list of \enquote{amazingly awesome hardware description language projects.}} All of them allow describing and/or generating hardware in more flexible and abstract ways than when using Verilog or VHDL.

Chisel has been proven to be a language that is capable of realizing large \acp{SoC} via the Rocket Chip project~\cite{Asanovic2016}. With the diplomacy framework, there is a tool that allows to negotiate parameters between individual Chisel generators~\cite{Cook2017} while staying entirely in the host language Scala, allowing for a highly integrated design process. There is very little literature on Chisel being used for small designs that perform low-level bit operations (see~\cite{Schoeberl2025} for a paper presenting initial ideas on smaller designs and building blocks and~\cite{Lennon2018} for a study similar to the one presented in this paper but targeting FPGAs).

The programming style for Chisel is, as in the host language Scala, a mixture between imperative and functional programming. The language Clash is fully embracing the functional approach: based on the pure functional programming language Haskell, it greatly raises the level of abstraction that can be used when describing hardware in it. According to the documentation~\cite{clash-tutorial}, Clash will not introduce any logic overhead:
\begin{quotation}
	What is always important to remember is that [...] the compiler will \emph{never} infer/invent more logic than what is specified in the circuit description. (emphasis theirs)
\end{quotation}
The high level of abstraction in Clash code, and to a certain extend in Chisel as well, may give rise to the impression that actually a \ac{HLS} is carried out. This is not the case. Consequently, this paper does \emph{not} deal with the issue of overhead introduced by \ac{HLS} with respect to RTL.

Although Clash is not as widely used as Chisel, it has seen some interest in implementing processors in it. There are fully verified RISC-V implementations~\cite{LionRiscV} and even a textbook on retrocomputing that re-implements multiple classic processors~\cite{Erdi2021}. 

The authors are not aware of any publications describing the usage of Clash to implement small designs using manual low-level optimizations.

The design used as a working example in this paper is an unsigned integer bit-serial \ac{MAC} unit written in Clash. The aim is to implement it as area-efficient as possible. This paper will investigate how different levels of abstraction will impact the size of the synthesized design and will also compare these findings against hand-written Verilog versions of the \ac{MAC} to investigate the impact of the \ac{HDL} used.

\section{The Clash HDL}
Clash is a functional programming language for circuit design based on the functional programming language Haskell.
It models combinatorial circuits as pure functions of type $f: A \rightarrow B$ where $A$ and $B$ are finite sets. The corresponding Clash notation is \mic{f :: A  -> B}. Very few restrictions apply in order for Clash to be able to synthesize code into either Verilog, SystemVerilog or VHDL; mainly that every data type needs to be finite and that recursions are statically known to terminate.

A simple implementation of computing the parity bit of a vector of \mic{n} bits in Clash is given by
\begin{minted}{Haskell}
	parityBit :: Vec n Bit -> Bit
	parityBit v = fold xor v
\end{minted}
The \mic{fold} function creates a tree structure of \mic{xor} operations of logarithmic depths in order to minimize the delay. One can see that Clash uses a high level of abstraction that allows the developer to clearly express the intent and, hence, focus on the design and not technical details.

Sequential circuits are modeled using the data type \mic{Signal dom a}. It represents an infinite stream of values of type \mic{a} in some clock domain \mic{dom}.\footnote{Concrete types in Clash have to start with a capital letter. A small letter indicates a placeholder for types, similar to generics in Scala.} When using this data type, Clash will generate the required registers to store the values. Clash will also generate clock, reset and enable signals based on the domain \mic{dom}. When working with multiple clock domains, Clash effectively prevents accidental clock domain crossings.

Clash comes bundled with data types that cover the common use cases: bit-vectors of arbitrary size $n$ (\mic{BitVector n}), (Un-)Signed integers of bit-width $n$ (\mic{Unsigned n}/\mic{Signed n}) as well as vectors of length $n$ storing arbitrary data of type \mic{a} (\mic{Vec n a}). Values that come with a valid status can be modeled via the \mic{Maybe a} data type where \mic{Just 4} is a valid integer and \mic{Nothing} indicates the absence of a valid value. This data type prevents the accidental use of bits that do not contain valid data.

Clash is very strict with respect to type checking. Even if two types have the same bit-representation, they can not be assigned to each other. The designer has to explicitly cast between the types. 

As the types are checked at compile time, no logic overhead is generated in the Verilog or VHDL files.

There are two main approaches when implementing hardware in Clash as discussed in the following two sections.

\subsection{Mealy Machines}

One method for implementing sequential circuits is to model them as a \emph{Mealy Machine}~\cite{Mealy1955}. This classical approach makes use two functions $T$ and $G$, the transition and output function, respectively. In Clash, a coalesced function that combines both functions in one is employed instead. Given a function $f$ of type \mic{f :: s -> i -> (s,o)} mapping the current state of type \mic{s} and the current input of type \mic{i} to a tuple consisting of the new state and the Mealy Machine's output of type \mic{o}. Clash provides the \mic{mealy} function of type
\begin{minted}{Haskell}
	mealy :: (s -> i -> (s, o)) -> s -> Signal dom i -> Signal dom o
\end{minted}
 that transforms such a coalesced transition/output function \mic{f} together with the initial state to a circuit implementing the Mealy machine. As an example consider the following Mealy machine that detect runs of consecutive \texttt{1}'s of length greater than three:\footnote{All shown codes examples are simplified or/and have parts omitted for didactic purposes. The full code of the presented MAC is available on GitHub.} 
\begin{minted}{Haskell}
detectRun :: Int -> Bit -> (Int, Bool)	
detectRun count b =
	if b == 1 
		then (count + 1, count + 1 > 3)
		else (0, False)

circuit :: Signal dom Int -> Signal dom Bool
circuit = mealy detectRun 0
\end{minted}
No manual definition of registers that store the internal state is necessary. Note that \mic{f} is a purely combinatorial function/circuit.

\subsection{State Monad-based Hardware}
Being a functional programming language, Haskell and, by extension, Clash, does not have a native concept of \emph{mutable state}. This is why the state was explicitly passed to the Mealy machine in the previous section. To avoid this and to write code that looks like the \enquote{conventional} imperative programming style, Haskell/Clash makes use of the so-called \emph{state monad} and provides a corresponding \mic{mealyS} function. 

We omit the technical details here and only point out that using a state monad-based programming approach allows describing hardware in a way that often more closely resembles the algorithmic description of the algorithm the hardware should implement, making the code easier to understand and maintain.

A state monad-based implementation of the previous example is given by the following code.
\begin{minted}{Haskell}
detectRunMonad :: Int -> State Int Bool
detectRunMonad i = do
  if i == 1
    then (modify' (+1))
    else (put 0)
  cnt <- get
  return $ cnt > 3

circuitMonad :: Signal dom Int -> Signal dom Bool
circuitMonad = mealyS detectRunMonad 0
\end{minted}

\section{Sequential Arithmetic}
\label{sec:mult}

Digital multipliers are well-studied circuits with a great amount of variation in designs and implementations seeking to optimize for one or more of area, speed, or power-consumption \cite{Ercegovac2003}. Given an unsigned $n$-bit multiplicand $X$, and an unsigned $m$-bit multiplier $Y$, their product is given by as
\begin{equation}
    X\cdot Y=\left(\sum_{i=0}^{n-1}x_i2^i\right)\cdot \left(\sum_{j=0}^{m-1}y_j2^j\right)=\sum_{i=0}^{n-1}\sum_{j=0}^{n-1}x_i\cdot y_j \cdot 2^{i+j}.
\end{equation}
The bit-wise product $x_i\cdot y_j$, is realized by a logic and-operation, while the scaling $2^{i+j}$ is a matter of bit-shifting or multiplexing. In a low-resource implementation, registers are used to hold each of the factors and the partial product. A single full-adder is used to iteratively calculate the partial sum of a single bit of the partial product, a bit-wise product of one bit from each operand, and the carry from the previous iteration.

This is illustrated in Algorithm~\ref{algo:fsmult}, where the outer loop ranges over the bits of $Y$ (we refer to this as a \enquote{round}) and the inner loop ranges over the bits of $X$. The bit-wise product $x_i\cdot y_j$ is computed by an and-operation. The new value of the current product bit is then the sum of the bit-wise product, the current product bit and the incoming carry.

After each round, the carry is propagated to the next product bit; which is possible because that bit is known to be zero. This is done by the \texttt{endRound} method (line~\ref{algo:eolStart}). The calls to \texttt{advanceX/Y} (lines~\ref{algo:stepEnd} and~\ref{algo:advanceY}) will be explained in Section~\ref{sec:access}. For now, consider them to be no-operations.

An exemplary calculation for $X=6$ and $Y=3$ is presented in the left part of Table~\ref{tab:example}. 

The \enquote{moving parts}, i.e., steps that render themselves amendable for design choices, will be discussed in the next section. When applicable, those moving parts are highlighted in blue in Algorithm~\ref{algo:fsmult}.

The sequential accumulating step following the sequential multiplication does not introduce further optimization potential and, hence, is omitted for readability.

\begin{algorithm}
	\DontPrintSemicolon
	\caption{Abstract view on the sequential multiplication.}
	\label{algo:fsmult}
	\KwData{Multiplicand and multiplier $x,y$ of width $n$ and $m$, respectively}
	
	\medskip
	$carry, product\gets 0$\;
	
	\medskip
	\tcp{$m$ partial products/rounds}
	\For{$roundCtr \gets 0$ \KwTo $m-1$}{\label{algo:rIdx}
		
		\tcp{$n$ bit positions within $X$}
		\For{$xCtr \gets 0$ \KwTo $n-1$} {\label{algo:xIdx}
			\medskip
			\tcp{The arithmetic part}
			$a \gets \textrm{\bfseries\color{blue}getXBit}(xCtr) \wedge \textrm{\bfseries\color{blue}getYBit}(roundCtr)$\;
			
			$p \gets \textrm{\color{blue}\bfseries getProdBit}(roundCtr, xCtr)$\; \label{algo:getPBit}

			$(carry,s) \gets \textrm{FullAdder}(a, p, carry)$\;
			
			\medskip
			\tcp{Updating the product}
			$\textrm{\bfseries\color{blue}setProdBit}(roundCtr, xCtr, s)$\; \label{algo:pSum}

			\medskip
			$\textrm{\bfseries\color{blue}advanceX}()$; \label{algo:stepEnd}
			
		}
		$\textrm{\bfseries\color{blue}endRound}(carry, xCtr, roundCtr)$\; \label{algo:eolStart}

		\smallskip
		$\textrm{\bfseries\color{blue}advanceY}()$\; \label{algo:advanceY}
		$carry \gets 0$~~\tcp*[f]{Reset carry}\; \label{algo:eolEnd}
	}
	\Return{product}\;
\end{algorithm}

\begin{table}
	\centering
\begin{tabular}{c|@{\hskip .3em}*{6}{c}@{\hskip .3em}|@{\hskip .3em}*{6}{c}}
	\toprule
	& \multicolumn{6}{c}{index-based} & \multicolumn{6}{c}{rotation-based} \\[.2em]
	~~~~~ & $x$ & $y$ & $c_{in}$ & $product$ & $c_{out}$ & $product'$ & $x$ & $y$ & $c_{in}$ & $product$ & $c_{out}$ & $product'$\\ \midrule
	P & 11\underline{0} & 1\underline{1} & 0 & 0000\underline{0} & 0 & 00000 & 
		11\underline{\textbf{0}} & 1\underline{1} & 0 & 0000\underline{\textbf{0}} & 0 & 00000\\

	P & 11\underline{0} & 1\underline{1} & 0 & 000\underline{0}0 & 0 & 00010 & 
		\textbf{0}1\underline{1} & 1\underline{1} & 0 & \textbf{0}000\underline{0} & 0 & 00001\\

	P & \underline{1}10 & 1\underline{1} & 0 & 00\underline{0}10 & 0 & 00110 & 
		1\textbf{0}\underline{1} & 1\underline{1} & 0 & 1\textbf{0}00\underline{0} & 0 & 10001\\[.35em]

	E & & & 0 & 0\underline{0}110 & 0 & 00110 & 
		 & & 0 & 11\textbf{0}0\underline{0} & 0 & 11000\\[.35em]

	P & 11\underline{0} & \underline{1}1 & 0 & 001\underline{1}0 & 0 & 00110 & 
	11\underline{\textbf{0}} & 0\underline{1} & 0 & \textbf{0}001\underline{1} & 0 & 00011\\

	P & 1\underline{1}0 & \underline{1}1 & 0 & 00\underline{1}10 & 1 & 00010 & 
		\textbf{0}1\underline{1} & 0\underline{1} & 0 & 1\textbf{0}00\underline{1} & 1 & 10000\\

	P & \underline{1}10 & \underline{1}1 & 1 & 0\underline{0}010 & 1 & 00010 & 
		1\textbf{0}\underline{1} & 0\underline{1} & 1 & 01\textbf{0}0\underline{0} & 1 & 01000\\[.35em]

	E &  &  & 1 & \underline{0}0010 & 1 & 10010 & 
		 & & 1 & 001\textbf{0}\underline{0} & 1 & 00101\\[.35em]
	E' &  &  &  & 00010 &  &  & 
		 & & 1 & 1001\underline{\textbf{0}} &  & \\
	\bottomrule
\end{tabular}
\caption{The computation carried out by Algorithm~\ref{algo:fsmult} for $X=6=110_2$ and $Y=3=11_2$ for the conventional, indexing-based data access (left) and the rotation-based data access (right). The bits used in each step are underlined. To help to understand the rotation-based method, the least-significant bits of $x$ and the product are highlighted in bold. The type of operation carried out in each row corresponds to the algorithm as follows: \textbf{P}roduct step (lines~\ref{algo:xIdx}--\ref{algo:stepEnd}) and \textbf{E}nd of round step (lines~\ref{algo:eolStart}--\ref{algo:eolEnd}).}
\label{tab:example}
\end{table}

\section{Implementation}
\label{sec:implementation}
The \ac{MAC}'s implementation features a $n$-bit and an $m$-bit input for the operands, a $n+m$-bit input for manually (pre-)setting the accumulator's value and enable bits for initiating a \ac{MAC} operation or setting the accumulator, respectively. The design outputs two $n+m$-bit outputs for the last product and the accumulator with corresponding valid bits.

The rest of this section describes the four design choices (\enquote{moving parts}) investigated in this paper.

\subsection{Mealy Machine vs. State Monad}
This design choice is the most abstract one: should the designer opt for the well-known approach of implementing a Mealy machine or should a more abstract approach be followed.

Listing~\ref{clash:mealy} shows a simplified code excerpt of how the corresponding Mealy machine could be implemented in Clash. Each state-input pair must be matched  and then used to compute the next state and the output.

\begin{listing}
\begin{minted}[xleftmargin=15pt, linenos, autogobble, tabsize=2,ignorelexererrors=true]{Haskell}
macMealy st@State{stage=Ready} Input{Nothing, Nothing} 
	= (st, getOutputs st)
macMealy st@State{stage=Ready} Input{Nothing, Just newAcc} 
	= (st', getOutputs st')
		where st ' = st{accumulator=newAcc}
macMealy st@State{stage=Multiplying, ..} Input{Nothing, Nothing} 
	= (st', getOutputs st')
		where st ' = --
-- more cases following (omitted here)
\end{minted}
\caption{Clash code using a Mealy machine to implement the \ac{MAC}. Shown is an excerpt of the multiplication part.}
\label{clash:mealy}
\end{listing}

Contrast this with the code shown in Listing~\ref{clash:state}. The state monad-based approach allows organizing the code in a fashion that it resembles an imperative programming language. The conditional setting 
of the accumulator value in Line~\ref{clash:state:shouldSetAccumulator} 
and the general linear flow give the impression of an imperative programming language. The \ac{MAC}'s state is handled implicitly; nothing has to be passed on explicitly. This can lead to code that is easier to read and might come more natural to implement designs in.

\begin{listing}
	\begin{minted}[xleftmargin=15pt, linenos, autogobble, tabsize=2]{Haskell}
macMonad accumulateFun multiplyFun Input{values, newAcc} = do
-- conditionally set the accumulator to a new value
case newAcc of !\label{clash:state:shouldSetAccumulator}!
	(Just acc) -> modify' (\s -> s{accumulator=bitCoerce acc})
	Nothing -> pure ()

-- start a new multiplication or continue with current computation
case values of
	Just (x,y) -> modify' (startMulState x y)
	Nothing -> do
		stage <- gets stage
		case stage of
			Ready -> pure () -- do nothing
			Multiplying -> modify' multiplyFun
			Accumulating -> modify' accumulateFun
			
-- retrieve the current output
gets extractOuptut
	\end{minted}
	\caption{Excerpt of the clash code implementing the \ac{MAC} using the state monad-based approach.}
	\label{clash:state}
\end{listing}


We posit that using the state monad-based approach is more straight-forward for designers used to imperative languages and, hence, might lead to shorter design cycles.

\subsection{Data Type of the Product}
There are two obvious ways to store the product bits: Either as a \mic{BitVector n}, that is, $n$ consecutive bits or using the type \mic{Vec n Bit}; a vector that has $n$ entries of type \mic{Bit}. Although these types look very similar, \mic{Vec} is more general as it allows to store arbitrary data types, not just \mic{Bit}s, and comes with a richer \ac{API} that, e.g., provides many folding operations that generate logarithmic reduction trees. Both types should be reduced to the same Verilog type when using \mic{Vec n Bit}. Hence, we expect that there should be no significant difference in the resulting circuits allowing designers to choose the data type/\ac{API} that suits them best or is most descriptive.

\subsection{Data Type of the Counters}

Clash provides the type \mic{Index n} consisting of the integers $[0,n-1]$ represented in the usual binary encoding using $\lceil\log_2(n)\rceil$ bits. This is the natural high-level choice for the counters $roundCtr$ and $xCtr$ (see Algorithm~\ref{algo:fsmult}, lines~\ref{algo:rIdx} and~\ref{algo:xIdx}).

As incrementing an integer requires a (possibly specialized) adder, we implement another type of counter based on one-hot encoding. For this, a \mic{BitVector n} is used that starts with the least significant bit set to one and all other bits set to zero. Each time the counter is increased, the set bit is shifted left by one until it reaches the most significant bit. This implementation does not need any adder hardware but requires designers to implement the overflow logic on their own and, therefore, is a rather low-level approach; even though this can effectively be hidden from the end-user by providing utility functions. For larger bit widths, we expect this implementation to become too unwieldy as it requires $n$ bits instead of the $\lceil\log_2(n)\rceil$ bits used by \mic{Index}.

\subsection{Accessing and Updating Data}
\label{sec:access}

Accessing and updating the product and reading the current position within $X$ and $Y$ can be carried out in different ways. Algorithm~\ref{algo:fsmult} uses functions allow for a different implementation (see lines~\ref{algo:getPBit} and~\ref{algo:pSum}).

One approach is to take the counter values and compute an index to access the product, e.g., $product_{roundCtr+xCtr}$ accesses the currently used product bit.\footnote{For the one-hot encoding approach, the position of the set bit has to be translated into a number first.} This is a high-level approach to implementing the \ac{MAC} in an intuitive fashion that closely resembles programming with arrays in regular programming languages. We call this the \emph{index-based} approach.

Reading and writing to arbitrary positions within the product requires (de-)multiplexing. To avoid this, another approach is investigated in this paper: replacing arbitrary indexing by indexing to the least-significant bit of a shifted product, e.g., $x[3] \triangleq (x \gg 3)[0]$. To avoid having to support shifting all values between $1$ and $n-1$, we only shift by 1 and store the updated value. This approach requires special care of how the product register is rotated as it needs to be reset after each round. As the patterns for $X$ repeat, we rotate this value instead of shifting in zeros. At the end of a round, the product needs to be shifted back by $n-1$ to start the next round at the next-higher bit. 

Implementing Algorithm~\ref{algo:fsmult} this way yields very difficult to understand Clash code. Designers should only opt to implement a design this way if there is a clear benefit with respect to non-functional properties, such as the circuit's area and make sure to \enquote{hide} it using a reasonable function names. We call this approach \emph{rotation-based}.

The right part of Table~\ref{tab:example} illustrates this approach for the multiplication  of $X=6$ and $Y=3$.

These two approaches lead to the implementations of the functions highlighted in blue in Algorithm~\ref{algo:fsmult} as shown in Table~\ref{tab:impl}. The \texttt{endRound} step becomes quite complicated for the rotation-based approach as shown in Algorithm~\ref{algo:endRound}.

\begin{table}
	\centering
\begin{tabular}{c|@{\hskip .3em}c@{\hskip .4em}|@{\hskip .4em}c}
	operation & index-based & rotation-based \\ \midrule
getX/YBit & x/y[ctr] & x/y[0] \\
getProdBit & product[roundCtr+xCtr] & product[0] \\
setProdBit & product[roundCtr+xCtr]=s & product[0]=s \\
advanceX & -- & x = rotate x 1 \\
advanceY & -- & y = y $\gg$ 1\\
endRound & product[roundCtr+xCtr+1]=c & see Algorithm~\ref{algo:endRound} \\
\end{tabular}
\caption{Implementations of the \enquote{moving parts} depending on the choice of the accessing/modification method.}
\label{tab:impl}
\end{table}

\begin{algorithm}
	\DontPrintSemicolon
	\caption{The complex \texttt{endRound} step when using the rotation-based approach.}
	\label{algo:endRound}
	$product \gets \textrm{rotate} ~ product ~ 1$\;
	$product[0] \gets c$\;
	\uIf{after last round}{
		$product \gets \textrm{rotate} ~ product ~ 1$\;
	}
	\Else{
		$product \gets \textrm{rotateBackwards} ~ product ~ (n-1)$\;
	}
\end{algorithm}

\subsection{Overview of the Designs}

The choices described above give rise to $16$ different designs per choice of bit-width in total as the following decisions can be made:
\begin{itemize}
	\item Mealy-machine-based or state monad-based modeling
	\item Data type of the product: \mic{BitVector} or \mic{Vec}
	\item Data type of the counters: one-hot encoding or using an off-the-shelf \mic{Index}
	\item Method for updating the product: index-based or rotation-based
\end{itemize}

The code necessary to generate all designs is publicly available on GitHub\footnote{\url{https://github.com/keszocze/mac-clash-iwsbp26}}.

\subsection{Testing the Clash Designs}

Modern \acp{HDL} usually also come with sophisticated testing and/or verification frameworks. Chisel, for example started with the rather simple \texttt{chiseltest} framework which was later extended by ChiselVerify~\cite{Dobis2021} to support, amongst others, \ac{CRV}~\cite{Mehta2018}.

 While Clash supports the automatic generation of classical Verilog test benches, we agree with \citeauthor{Erdi2021}, quoting from~\cite[Page 24]{Erdi2021}:
 \begin{quote}
 	\enquote{One common trait among FPGA vendor tools is that they range from horribly painful to painfully horrible. In this book, interaction with the FPGA toolchain is restricted to a minimum by writing everything we can in Clash}
 \end{quote}

Clash designs are valid Haskell code and, therefore, can be tested using any of the many software testing frameworks available for Haskell.\footnote{The authors are fully aware that this only covers functional testing and that it relies on the assumption of the Clash compiler being bug-free.} The program carrying out the tests is then similar to running a Verilator\footnote{\url{https://www.veripool.org/verilator/}} simulation.

A particularly interesting type of testing is \emph{property-based testing}~\cite{Fink1997,Goldstein2024}. This approach differs from classical unit testing in that it does not rely on pre-specified inputs and the corresponding golden answers but allows to specify properties (such as $\forall x, y: \texttt{MAC}(x,y, acc) = x\cdot y + acc$) 
 that are automatically validated by generating random values and checking whether the property holds. 

To prune the design space, we only investigate design with $n=m$.
For bit-widths $n > 8$, the designs have been tested with 200 randomly generated inputs. For bit-widths $n$ with $n \le 8$, all possible $2^{n+n}$ inputs have been simulated, effectively fully verifying the smaller designs. We do expect the test results to also hold for cases where $n\ne m$.

\section{Experimental Evaluation}

\subsection{Verilog Designs}
We also implemented the MAC unit in Verilog in order to investigate whether differences in the area of synthesized designs stem from different implementation choices or the used \ac{HDL}. We are further interested to see whether there is a general overhead introduced by either of the languages.

The architectural variants -- indexing vs. rotating and base-2 counters vs. one-hot counters -- were also directly implemented in Verilog code. The interface’s timing behavior was matched between the Clash and Verilog implementations, and was verified using simulations of the hand-written Verilog code together with the Verilog code compiled from the Clash implementation. To an external observer, all implementations of the circuit would be identical.

\subsection{Experimental Setup}

Both Clash-derived and hand-written Verilog code was passed through the initial steps of the default LibreLane flow for chip design using the SkyWater  \SI{130}{nm} PDK, including linting, error-checking, and synthesis using the open-source tool Yosys. The synthesis step produces a netlist of standard cells defined in the process design kit (PDK). Standard cells correspond to memory elements in the form of D-flipflops and logic elements in the form of logic gates such as \enquote{nand} and \enquote{nor} as well as slightly more advanced logic operations on a few bits. The PDK describes the function of the standard cells as well as their individual area on a chip. For comparing the various implementations, the total area of the synthesized circuit as well as the area of the sequential (memory) and logic parts is considered individually.

The decision to stop after synthesis was made deliberately. Many additional steps are needed to produce a functioning chip, but none of these steps take the Clash-derived or hand-written Verilog files as input. They are rather based off the synthesized netlist or other representations derived from the netlist. Each of these steps may introduce changes to the circuits’ areas due to the physical constraints of semiconductors and the manufacturing steps, but as these changes are not directly related to the initial description of the circuit, they are considered confounding factors better left out of the comparison.

We ran the flow for the 27 bit-widths $n=2,3,\ldots,16,20,24,\ldots, 64$ and all $16$ configurations as described in
Section~\ref{sec:implementation}, effectively performing a full \ac{DSE}.

\subsection{Evaluating the Design Choices}

In this section, we will discuss the relative difference between circuits varying in only one aspect, i.e. a single design choice (see Section~\ref{sec:implementation}). More concretely, given two different versions $v_1$ and $v_2$ of the MAC unit, we compute $$\frac{\textrm{size}(v_1) - \textrm{size}(v_2)}{\textrm{mean}(\textrm{size}(v_1), \textrm{size}(v_2))},$$ i.e. the difference of all sizes for each design choice divided by the mean over all sizes. 
The relative difference is shown as a function of the bit-width.

\subsubsection{Variations within the Clash Designs}

\begin{figure}[t]
	\centering
        \includegraphics[width=\linewidth]{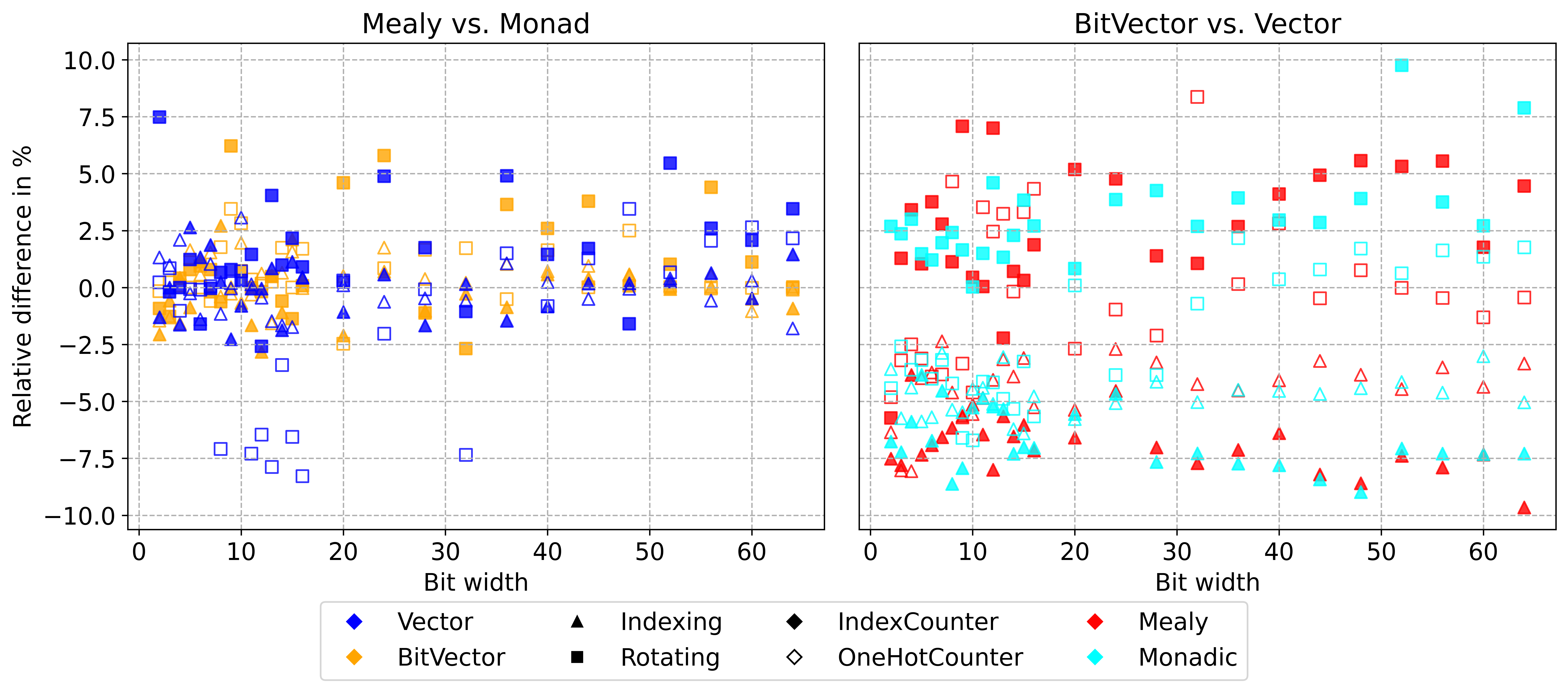}
	\caption{Clash design choices with low influence on circuit sizes, left: Mealy machine vs.\ state monfad, right \mic{BitVector} vs.\ \mic{Vec}. Positive values indicate larger area for Mealy (left) and BitVector (right) based implementations. The legend shows individual factors by color, shape, and filled/hollow markers.}
        \label{fig:clash_low_impact}
\end{figure}

Starting with the language specific variations of the Clash implementations, specifically expressing the circuits as a Mealy machine as opposed to a state monad (see Fig.~\ref{fig:clash_low_impact}, left) and using the \mic{BitVector} type as opposed to a \mic{Vector} of bits type, minimal impact is seen on the circuits’ areas (see Fig.~\ref{fig:clash_low_impact}, right). This indicates that little to no overhead is associated with either of these variations. Consequently, for the investigated circuits, designers can use any of these approaches with little concern for the size of the synthesized circuit.

\begin{figure}[t]
	\centering
        \includegraphics[width=\linewidth]{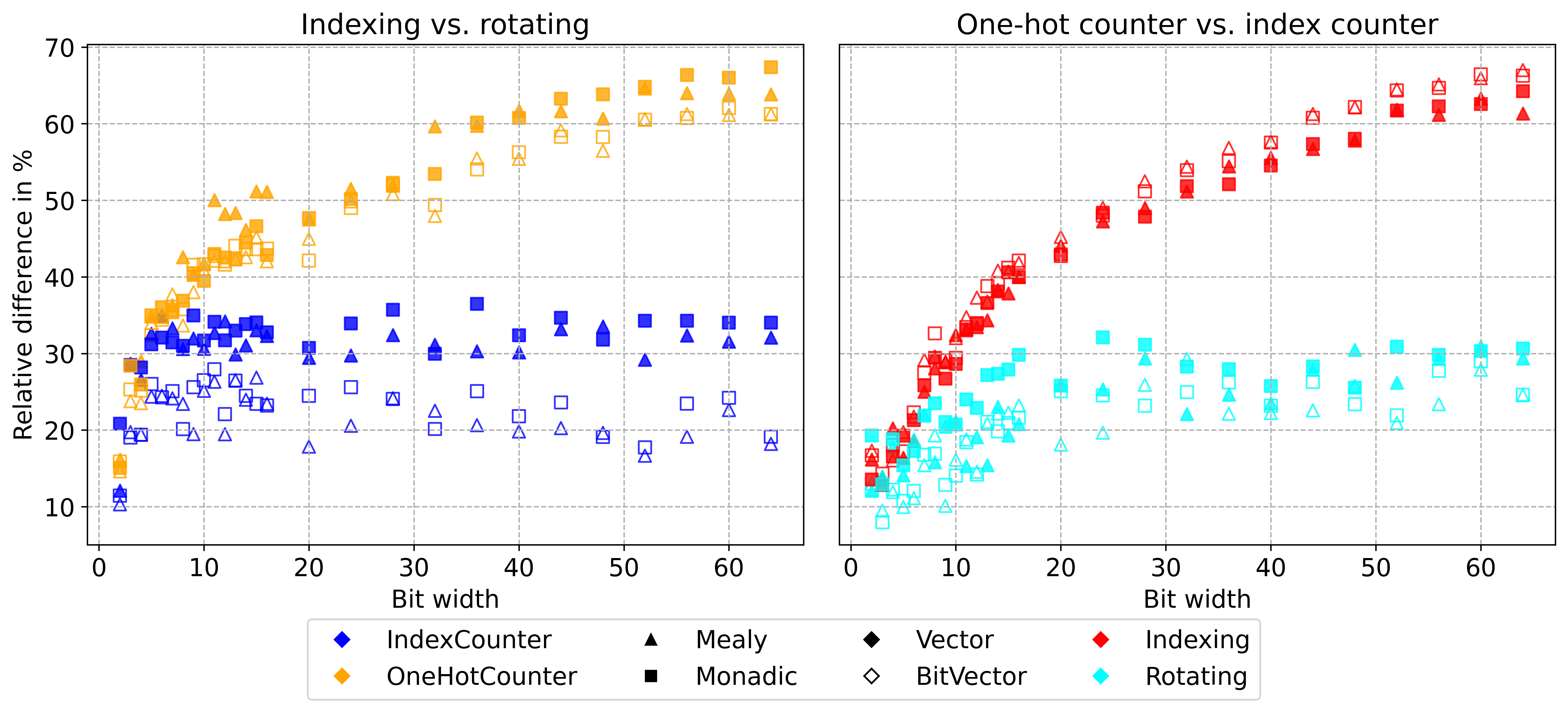}
	\caption{Clash design choices with strong influence on circuit sizes, left: Indexing vs.\ rotating accesses, right: One-hot vs.\ index counters. Positive values indicate larger area for indexing (left) and one-hot counter (right) based implementations.}
        \label{fig:clash_high_impact}
\end{figure}

Moving into the architectural variations, comparing indexing into vs. rotating the registers, a clear difference is seen (see Fig.~\ref{fig:clash_high_impact}, left). This is to be expected, as N-input multiplexers, required for indexing, typically are implemented as a tree of 2- or 3-input multiplexers giving an area scaling of $\mathcal{O}(N\log_2(N))$. This should be compared to the variant with rotating (barrel-shift) registers that only requires $\mathcal{O}(1)$-sized logic to control the shifting. The results show this trend. Two groupings are seen corresponding to the other architectural variation, where one grouping corresponds to the variation using base-2 counters, and the other grouping corresponds to using one-hot counters. Considering the sequential and combinatorial (logic) areas, zero difference is seen in the sequential areas indicating that exactly the same number of bits is stored in registers for both multiplexing and rotating, which is to be expected, whereas the difference seen in total area is matched by the difference in logic area, which includes multiplexers.

Considering the comparison between counter types, again a clear difference is seen (see Fig.~\ref{fig:clash_high_impact}, right). Using base-2 counters (i.e., \mic{Index}) requires registers of $\lceil \log_2(N)\rceil$ bits and having adders of the same width, whereas using one-hot counters requires a shift-register and $\mathcal{O}(1)$ logic to control this register. Looking at the area of sequential standard cells (flipflops), this difference in register width is clearly seen and it is identical for all variations (all series have near-identical relative differences). Looking at the difference in logic area, again two groupings are seen corresponding to indexing vs. rotating. The circuits with base-2 counters have lower area than those with one-hot counters, and for the first grouping corresponding to rotating the difference is roughly constant, whereas it depends on the bit-width for the second grouping corresponding to multiplexing. This is as expected.

\subsubsection{Variations within the Verilog Designs} 
\begin{figure}[t]
  \centering
  \includegraphics[width=\linewidth]{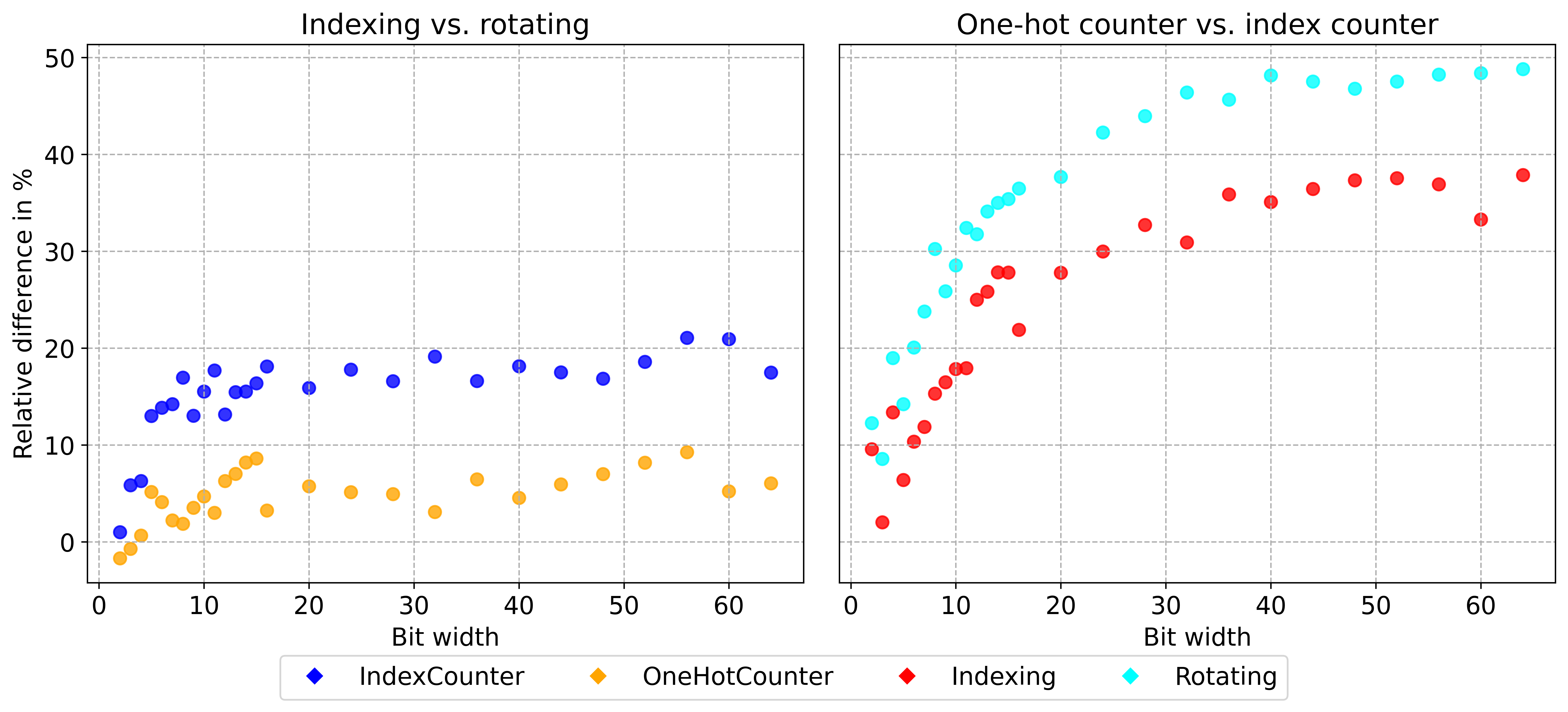}
  \caption{Design choices in Verilog with a large influence on the resulting circuit size, left: indexing vs.\ rotating accesses, right: One-hot vs.\ index counters. Positive values indicate larger area for indexing (left) and one-hot counter (right) based implementations.}
  \label{fig:verilog}
\end{figure}

Comparing architectural variations in the Verilog implementations shows similar trends as for the implementations written in Clash: Indexing takes up larger area than rotating, albeit less pronounced than in the Clash case (see Fig.~\ref{fig:verilog}, left), and one-hot counters take up larger area than base-2 counters (see Fig.~\ref{fig:verilog}, right) with all the difference being in the logic for multiplexing vs. rotating, and the counter variations having both different sequential and logic areas, as expected.

\subsubsection{Clash vs. Verilog}
Since little variation was seen between the Mealy machine and state monad-based approach and between using either \mic{BitVector} or \mic{Vector} for the Clash implementations, the hand-written Verilog is compared to the Clash-implementations using a Mealy machine and the \mic{Vector} type.

\begin{figure}
	\centering
        \includegraphics[width=0.9\linewidth]{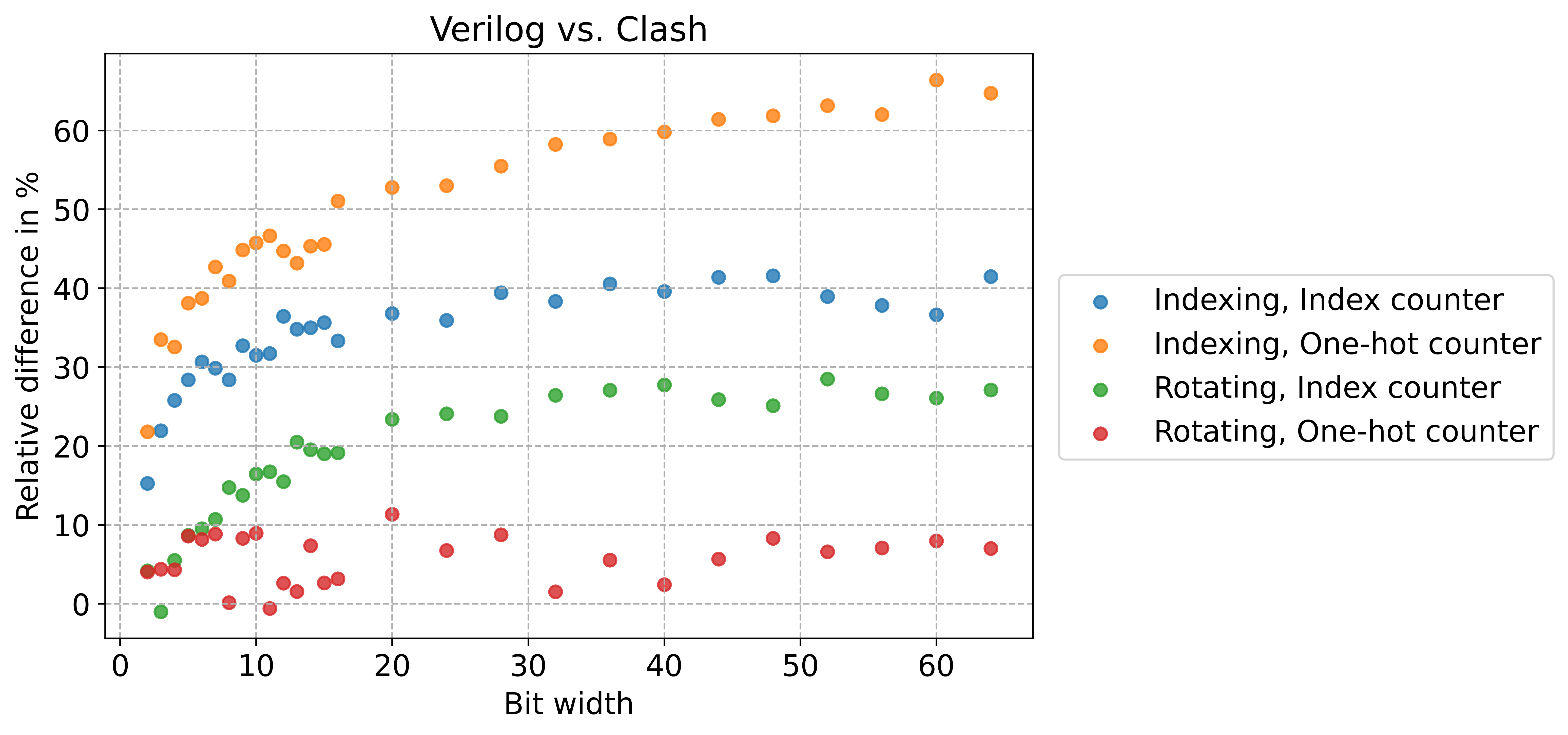}

	\caption{Size comparison of hand-written Verilog to Clash. Positive values indicate larger area for Clash-implementations.}
	\label{fig:clash_vs_verilog}
\end{figure}

The area of the hand-written Verilog circuits is lower than that of the Clash implementations in practically all cases (see Fig.~\ref{fig:clash_vs_verilog}). The difference, however, varies across the four architectural variations. The largest difference is seen for the circuits using multiplexing. As previously described, all circuits have identical behavior on the interface, therefore it can be concluded that the hand-written Verilog implementations produce netlists of fewer or smaller standard cells than the Clash implementations. This corresponds to a reduced logic complexity, and while the details in why these differences arise are yet to be investigated, it is a strong indication that either the Clash implementation or the Clash compiler results in higher complexity for the same operation than a direct implementation in a lower-level language such as Verilog. Interestingly, both for the area of sequential and logic cells, a dependence is seen on the bit-width with Verilog requiring increasingly less area of both types with increasing bit-width. This indicates that the Clash implementations both have a larger state and a higher logic complexity.

\section{Conclusion \& Outlook}
While the results shown in the previous section indicate smaller state and lower logic complexity of hand-written Verilog compared to the Clash implementations, other factors not included in the present study need to be considered before making a conclusion about the benefits of one language over another. For one, while area is an important aspect of chip design, other aspects such as latency, throughput, power, and energy consumption are also highly important metrics. For a study as the one presented here with more than a thousand synthesized circuits, the area of the synthesized netlist is the simplest to extract, and given the stated differences in state size and logic complexity, it would be highly unexpected to see opposing trends for area, latency, throughput, or power consumption.

However, these \enquote{hard} metrics are not always the most important ones in the real world. For example, the chip design tools tend to create a 2-D bounding box containing the interface on the periphery and the functionality (standard cells) inside the box. For simple circuits with a \enquote{large} interface - measured as the total number of bits -- the logic complexity may be of secondary importance as the size of the bounding box will be decided by the interface size, while the circuit may only take up a small fraction of the box’s area. In such cases, the agility in development cycles offered by higher-level languages may be preferable over having closer-to-optimal designs from the start. It does not seem farfetched to envision a design flow starting from descriptions using high-level languages in an agile workflow followed by profiling and targeted low-level optimizations or even re-implementations in low-level languages where needed. Further study is needed to evaluate the efficacy of such design flows.

As additional future work, we intend to extend our comparison to the Chisel \ac{HDL} (similar to \cite{Lennon2018}) and Amaranth. We will also try to match the areas obtained from the hand-written Verilog code to see how much effort is necessary to become truly competitive with respect to area. We also intend to compare the result obtained by the open-source tools to commercial tools. 

\printbibliography
\end{document}